\documentclass{emulateapj}
\begin{document}
\def\hh{\, h^{-1}}
\newcommand{\wth}{$w(\theta)$}
\newcommand{\xir}{$\xi(r)$}
\newcommand{\Lya}{Ly$\alpha$}
\newcommand{\Lyb}{Lyman~$\beta$}
\newcommand{\Hb}{H$\beta$}
\newcommand{\HI}{H{\sc I}}
\newcommand{\Msun}{M$_{\odot}$}
\newcommand{\sfr}{M$_{\odot}$ yr$^{-1}$}
\newcommand{\sfrd}{M$_{\odot}$ yr$^{-1}$ Mpc$^{-3}$}
\newcommand{\cld}{erg s$^{-1}$ Hz$^{-1}$ Mpc$^{-3}$}
\newcommand{\dnsty}{$h^{-3}$Mpc$^3$}
\newcommand{\za}{$z_{\rm abs}$}
\newcommand{\ze}{$z_{\rm em}$}
\newcommand{\cmtwo}{cm$^{-2}$}
\newcommand{\nhi}{$N$(H$^0$)}
\newcommand{\degpoint}{\mbox{$^\circ\mskip-7.0mu \,$$~$}}
\newcommand{\halpha}{\mbox{H$\alpha$}}
\newcommand{\hbeta}{\mbox{H$\beta$}}
\newcommand{\hgamma}{\mbox{H$\gamma$}}
\newcommand{\kms}{\,km~s$^{-1}$}      % note leading thinspace
\newcommand{\minpoint}{\mbox{$'\mskip-4.7mu \mskip0.8mu \,$$~$}}
\newcommand{\mv}{\mbox{$m_{_V}$}}
\newcommand{\Mv}{\mbox{$M_{_V}$}}
\newcommand{\Mh}{M_h}
\newcommand{\peryr}{\mbox{$\>\rm yr^{-1}$}}
\newcommand{\secpoint}{\mbox{$''\mskip-7.6mu \\,$}}
\newcommand{\sqdeg}{\mbox{${\rm deg}^2$}}
\newcommand{\squig}{\sim\!\!}
\newcommand{\subsun}{\mbox{$_{\twelvesy\odot}$}}
\newcommand{\et}{{\it et al.}~}
\newcommand{\er}[2]{$_{-#1}^{+#2}$}
\def\h50{\, h_{50}^{-1}}
\def\hbl{km~s$^{-1}$~Mpc$^{-1}$}
\newcommand{\lseq}{\mbox{\raisebox{-0.7ex}{$\;\stackrel{<}{\sim}\;$}}}
\newcommand{\gseq}{\mbox{\raisebox{-0.7ex}{$\;\stackrel{>}{\sim}\;$}}}
\newcommand{\fesc}{f_{\rm esc}}
\newcommand{\fescrel}{f_{\rm esc,rel}}
\def\spose#1{\hbox to 0pt{#1\hss}}
\def\simlt{\mathrel{\spose{\lower 3pt\hbox{$\mathchar"218$}}
     \raise 2.0pt\hbox{$\mathchar"13C$}}}
\def\simgt{\mathrel{\spose{\lower 3pt\hbox{$\mathchar"218$}}
     \raise 2.0pt\hbox{$\mathchar"13E$}}}
\def\arcs{$''~$}
\def\arcm{$'~$}
\def\see{\mbox{$^{\prime\prime}$}}
\newcommand{\wu}{$U_{300}$}
\newcommand{\wb}{$B_{435}$}
\newcommand{\wv}{$V_{606}$}
\newcommand{\wi}{$i_{775}$}
\newcommand{\wz}{$z_{850}$}
\newcommand{\hmpc}{$h^{-1}$Mpc}
\newcommand{\um}{$\mu$m}
\newcommand{\muref}{$\mu_{5}^{-1}~$}

\title{Characterizing faint galaxies in the reionization epoch: LBT confirms two
$L<0.2L^{\star}$ sources at $z=6.4$ behind the CLASH/Frontier Fields cluster MACS0717.5+3745}

\author{\sc E. Vanzella\altaffilmark{2},
            A. Fontana\altaffilmark{3},
            A. Zitrin\altaffilmark{4,$\dag$},
            D. Coe\altaffilmark{5},
            L. Bradley\altaffilmark{5},
            M. Postman\altaffilmark{5},
            A. Grazian\altaffilmark{3},
            M. Castellano\altaffilmark{3},
            L. Pentericci\altaffilmark{3},
            M. Giavalisco\altaffilmark{6},
            P. Rosati\altaffilmark{7},
            M. Nonino\altaffilmark{8},
            R. Smit\altaffilmark{9},
            I. Balestra\altaffilmark{8},
            R. Bouwens\altaffilmark{9},
            S. Cristiani\altaffilmark{8},
            E. Giallongo\altaffilmark{3},
            W. Zheng\altaffilmark{10},
            L. Infante\altaffilmark{11},
            F. Cusano\altaffilmark{2},
            R. Speziali\altaffilmark{3}
  }

\affil{$^{2}$INAF -- Bologna Astronomical Observatory, via Ranzani 1, I-40127 Bologna, Italy}
\affil{$^{3}$INAF - Rome Astronomical Observatory, Via Frascati 33, I-00040 Monteporzio Roma, Italy}
\affil{$^{4}$Cahill Center for Astronomy and Astrophysics, California Institute of Technology, MS 249-17, Pasadena, CA 91125, USA}
%\affil{$^{4}$College of General Education, Osaka Sangyo University, 3-1-1, Nakagaito, Daito, Osaka 574-8530, Japan}
%\affil{$^{5}$INAF - Rome Astronomical Observatory, Via Frascati 33, I-00040 Monteporzio Roma, Italy}
%\affil{$^{6}$NOAO, PO Box 26732, Tucson, AZ 85726, USA}
%\affil{$^{7}$Jet Propulsion Laboratory, California Institute of Technology, Mail Stop 169-527, Pasadena, CA 91109, USA}
\affil{$^{5}$ STScI, 3700 San Martin Dr., Baltimore, MD 21218, USA}
\affil{$^{6}$Department of Astronomy, University of Massachusetts, Amherst MA 01003, USA}
\affil{$^{7}$ Dipartimento di Fisica e Scienze della Terra, Universit\`a di Ferrara, via Saragat 1, 44122 Ferrara, Italy}
%\affil{$^{9}$Department of Astronomy, University of California, Berkeley, CA  94720, USA}
%\affil{$^{10}$ESO, Karl Schwarzschild Strasse 2, 85748, Garching, Germany}
%\affil{$^{4}$NOAO, PO Box 26732, Tucson, AZ 85726, USA}
%\affil{$^{8}$JPL, California Institute of Technology, Mail Stop 169-527, Pasadena, CA 91109}
\affil{$^{8}$ INFN, National Institute of Nuclear Physics, via Valerio 2,  I-34127 Trieste, ITALY}
\affil{$^{9}$Leiden Observatory, Leiden University, PO Box 9513, 2300 RA Leiden, The Netherlands}
\affil{$^{10}$Department of Physics and Astronomy, The Johns Hopkins University, 3400 North Charles Street,
Baltimore, MD 21218, USA}
\affil{$^{11}$ Institute of Astrophysics, Pontificia Universidad Catolica de Chile,
V. Mackenna 4860, 22 Santiago, Chile}
\affil{$^{\dag}$Hubble Fellow}

\altaffiltext{1}{The Large Binocular Telescope (LBT) is an international
collaboration among institutions in the United States,
Italy and Germany. LBT Corporation partners are: The University of Arizona
on behalf of the Arizona university system; Istituto Nazionale di Astrofisica, Italy;
LBT Beteiligungsgesellschaft, Germany, representing the Max-Planck Society,
the Astrophysical Institute Potsdam, and Heidelberg University; The Ohio State
University, and The Research Corporation, on behalf of The University of Notre Dame,
University of Minnesota, and University of Virginia.}

\begin{abstract} 
We report the LBT/MODS1 spectroscopic confirmation of two images 
of faint Lyman alpha emitters at $z=6.4$ behind the Frontier
Fields galaxy cluster MACSJ0717.5+3745. 
A wide range of lens models suggests that the two images 
are highly magnified, with a strong lower limit of $\mu>5$. 
These are the faintest $z > 6$ candidates spectroscopically confirmed to date.
These may be also multiple images of the same $z = 6.4$ source 
as supported by their similar intrinsic properties, but the lens 
models are inconclusive regarding this interpretation. To be cautious, 
we derive the physical properties of each image individually.
Thanks to the high magnification, the observed near-infrared (restframe 
ultraviolet) part of the spectral energy distributions and $Ly\alpha$ 
lines are well detected with $S/N(m_{1500})\gtrsim 10$ and $S/N(Ly\alpha)\simeq 10-15$.
Adopting $\mu > 5$, the absolute magnitudes, 
$M_{1500}$, and $Ly\alpha$ fluxes, are fainter than
$-18.7$ and $2.8\times10^{-18}~erg s^{-1} cm^{-2}$, respectively. We
find a very steep ultraviolet spectral slope $\beta=-3.0 \pm 0.5$
($F_{\lambda}=\lambda^{\beta}$), implying that these are very
young, dust-free and low metallicity objects, made of standard stellar
populations or even extremely metal poor stars (age $\lesssim 30$Myr,
E(B-V)=0 and metallicity $0.0 - 0.2 Z/Z_{\odot}$).  The objects are
compact ($< 1 kpc^{2}$), and with a stellar mass $M_{\star} < 10^{8}
M_{\odot}$. The very steep $\beta$, the presence of 
the $Ly\alpha$ line and the intrinsic FWHM ($<300~kms^{-1}$) of these newborn 
objects do not exclude a possible leakage of ionizing radiation. 
We discuss the possibility 
that such faint galaxies may resemble those responsible for cosmic 
reionization.

\end{abstract}
\keywords{dark ages, reionization, first stars --- cosmology: observations --- galaxies: formation}
%galaxies: evolution}

\section{Introduction}
The investigation of the distant Universe and the processes that led to the
reionization of the intergalactic medium, are amongst the major goals of
observational cosmology (Robertson et al. 2010). While there are
tens (a few) spectroscopic confirmations of galaxies at redshift 6(7)
(e.g., Vanzella et al. 2009, 2011), accessing the faint-luminosity regime
down to $\lesssim 0.2L^{\star}$ remains challenging even with 8-10m class telescopes,
especially for $z>6$.
Before the advent of next generation observatories like JWST and the
extremely large telescopes, the only viable way to pursue extremely faint distant
objects, and investigate
the nature of their stellar populations (even PopIII), is to exploit strong lensing
magnification (e.g., Zackrisson et al. 2012, 2013).
To this aim, Bradley et al. (2013) (B13, hereafter) selected magnified candidate galaxies at
redshift 6 -- 8 fully exploiting the 16-bands photometry of the CLASH survey
(Postman et al. 2011), and found agreement down to $\sim 27$ mag with the UV luminosity
functions of blank fields.
After the completion of the CLASH program, the investigation
of the high-z universe is now continuing with the ultradeep HST Frontier
Fields campaign (FF hereafter), that includes four CLASH galaxy
clusters\footnote{http://www.stsci.edu/hst/campaigns/frontier-fields/}.
%%%%%%%%%%%%%%%%%
\begin{figure*}
 \epsscale{0.9}
 \plotone{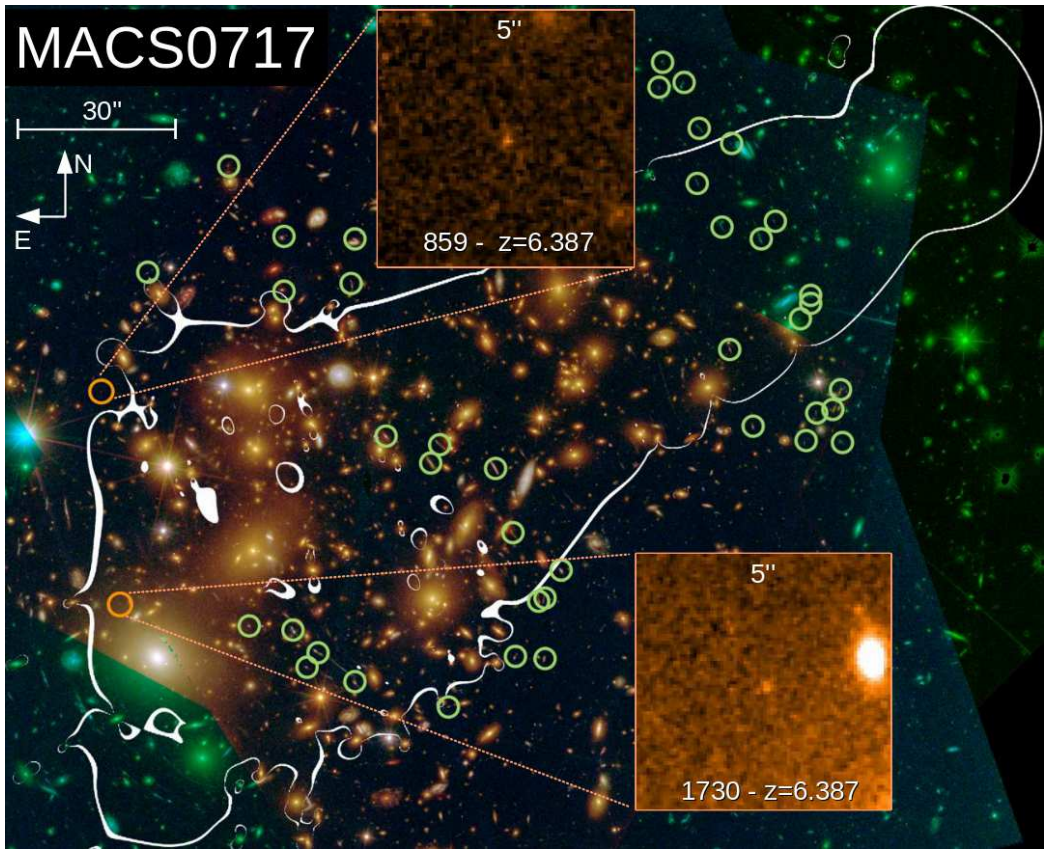} 
\caption{Figure shows the 16-band CLASH RGB false-color image of
MACSJ0717.5+3745, with the two $z=6.4$ spectroscopically confirmed images
marked with $\emph{red circles}$ (the insets show the J125 zoom).
The critical curves ($\mu>100$ here) for a source at
$z_{s}=6.4$ from the revised Zitrin et al. model are overlaid in
$\emph{white}$. The $\emph{green}$
circles mark the multiple images used as constraints (see Zitrin et al.
2009, Limousin et al. 2012, Medezinski et al. 2013). As can be seen, the
two $z=6.4$ objects lay (a) close to the critical curves, and (b) in
regions in which there are hardly other multiple images known, so that the
exact position of the critical curves is not perfectly constrained. The
proximity to the critical curves results in very high magnifications, of the
order of few to few dozen, and correspondingly, large errors on these
estimates. Still, all probed models (see \S 2.1) yield
$\mu>5$ for both images, which we have adopted throughout this work as our
lower limit, for conservative results. \label{pano}}
\end{figure*}
%%%%%%

Accessing the faint luminosity regime ($L\lesssim 0.2L^{\star}$) at $z>6$
is crucial in the context of cosmic reionization (e.g., Fontanot et al. 2013):
faint galaxies dominate the global ultraviolet luminosity density (Bouwens et al. 2007)
and possibly have an escape fraction of ionizing radiation larger than the
brighter counterparts (e.g., Ferrara \& Loeb 2013; Yajima et al. 2011).

Here we report on the LBT/MODS1 spectroscopic confirmation of two
faint $z=6.4$ sources, significantly magnified by the FF galaxy
cluster MACSJ0717.5+3745 (Ebeling et al. 2007), 
study their physical properties, and discuss
the contributions of such objects to the reionization of the IGM.

Throughout this paper a concordance $\Lambda$CDM cosmology with $\Omega_m=0.3$, 
$\Omega_{\Lambda}=0.7$ and $H_0=70$ \hbl~is adopted,
and magnitudes are in AB scale.

\subsection{Target selection and magnification}

%%%%%%%%%%%%%%%%%
\begin{figure*}
 \epsscale{1.0}
 \plotone{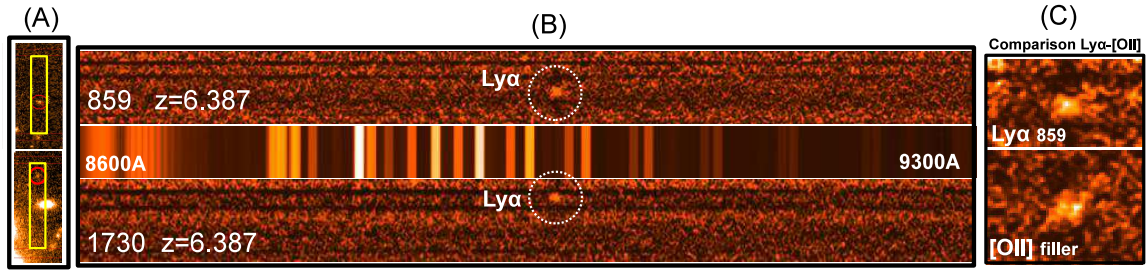}
\caption{
{\bf Panel (A):} the position of targets 859 (top) and 1730 (bottom) in the MODS1 slits over the J125 band are shown.
{\bf Panel (B):} the two dimensional spectra with the $Ly\alpha$ lines (marked with dotted circles) and the sky spectrum
 are shown. {\bf Panel (C):} The $Ly\alpha$ line of 859 compared with a
low-z [O\,{\sc ii}]3727 doublet (tilted) identified in the same mask (filler object) is shown. \label{Fig1}}
\end{figure*}
%%%%%%%%%%%%%%%%

B13 selected 15 magnified $z\simeq 6$ galaxy candidates behind the FF
galaxy cluster MACSJ0717.5+3745, by using their drop-out features and
corresponding photometric redshift estimate. We report
here the spectroscopic observations of two candidates from their
sample, macs0717\_0859 and macs0717\_1730 (859 and 1730 for short, hereafter), 
with photometric redshifts of $6.1 \pm 0.2$ and $6.0^{+0.2}_{-0.3}$, respectively.
The magnifications reported in B13 were $\mu=15.6$ (859) and $\mu>100$ (1730)
(i.e. the latter unconstrained since the object is too close to the critical curves).
The magnification estimates were based on the revised
lens model by Zitrin et al. (2009; see also Medezinski et al. 2013)
who first performed the strong-lensing analysis for this cluster,
uncovering that is the largest magnifying lens known to date (see Figure~\ref{pano}).
Here we have also estimated the magnifications from several other lens models
made for the Frontier Fields program (including a refurbished version
of the Zitrin et al. model used in B13), by running the Magnification
Calculator available online. \footnote{http://archive.stsci.edu/prepds/frontier/lensmodels/}
The estimate from different groups, methods and
assumptions span the range between 5 and 70, with some solutions even
higher than 100 within the 68\% confidence interval. The medians among
the different models are: $\mu=17.4^{+25}_{-13}(^{+50}_{-12})$ for 1730
and $\mu=6.9^{+1}_{-1}(^{+30}_{-2})$ for 859, where statistical and
systematic errors (in parentheses) are quoted.  The models for this
lens are still not fully constrained in the regions where the two
$z=6.4$ are detected, both due to proximity to the critical curves,
and, lack of multiple-images constraints nearby. 
We also acknowledge the possibility that the two sources presented here are
actually counter images of a single background galaxy, as some of the
models provided by the different groups predict counter images 
within few, to dozen arcseconds, from the
location of the other $z=6.4$ object. We did not detect, however, any
additional counter images where the models predict them (although possibly, 
due to lesser magnification where other images are predicted). 

As not all 
models predict counter images, and predicted counter images were not 
identified in the data, it cannot be unambiguously determined if indeed the 
two objects are images of the same source. What is relevant here, though, is the
agreement among the different models that the sources are strongly
magnified ($\mu>5$), and the single or double nature does not alter our
findings on the derived physical properties.
In the following, to be most conservative, we derive
rest-frame quantities by adopting $\mu=5$ for both sources, and express the
results in terms of  $\mu_5 = \mu /5$. 

\section{Data and sample selection}

\subsection{Spectroscopic observations with LBT/MODS1}
The spectroscopic observations have been performed in dual mode with the MODS1 instrument
at the LBT, that exploits the two red (5800-10300\AA) and blue (3200-6000\AA) channels,
yielding a total spectra coverage from 3200 to $\sim$10300\AA~on source. The red
G670L and blue G400L grisms with a slit width of $1''$ have been adopted,
providing a spectral resolution of $R\simeq 1500$ for both.
Science frames of 1200s have been acquired with a dithering pattern of $1.5''$ shift
along the slit for a total integration time of 16800s for 859, and 11200s for
1730. The average seeing conditions were $\simeq 1.0''$. Data reduction has been performed
with the MODS1 spectroscopic reduction pipeline based on VIPGI tasks
(Scodeggio et al. 2005).\footnote{{\it http://lbt-spectro.iasf-milano.inaf.it/pipelinesInfo/}}
In the two slits located on 859 and 1730, two emission lines are clearly detected at 8980\AA~and
8981\AA, respectively, with observed fluxes of $1.4 \times 10^{-17} erg s^{-1} cm^{-2}$
(with $S/N = 15$) and $\simeq 1.0 \times 10^{-17} erg s^{-1} cm^{-2}$ (with $S/N = 9$),
respectively (see Figures~\ref{Fig1} and ~\ref{Fig2}).

\section{Results}

%%%%%%%%%%%%%%%%%
\begin{figure}
 \epsscale{1.0}
 \plotone{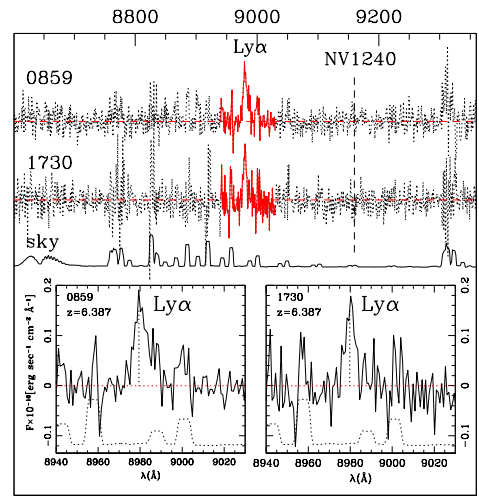}
\caption{One dimensional spectra of 859 and 1730 (top dotted). The highlighted red parts of the spectra
are zoomed in the bottom panels with the sky spectrum (dotted). The position of the N\,{\sc V}$\lambda1240$ 
line is also shown with a vertical dashed line. \label{Fig2}}
\end{figure}
%%%%%%

$\bullet$~{\it Nature of the lines:}
The large spectral coverage ($3200-10300$\AA) allows us to exclude low redshift solutions like
$H\alpha$ at $z=0.37$ or [O\,{\sc iii}]$\lambda 5007$ at $z=0.79$, that would be in contrast with the
single line detection. The only possible degeneracy is among
[O\,{\sc ii}]$\lambda 3727$ and $Ly\alpha$. However, [O\,{\sc iii}]$\lambda 3727$ can be reliably excluded because of
the following reasons:
(1) the doublet [O\,{\sc ii}]$\lambda 3726-3729$ is resolved in the present observations
(see an example in Figure~\ref{Fig1}, panel C) and
(2) the observed equivalent width (see below) of the lines is not compatible with
the typical values observed at $z<1.5$, i.e., they are too large
(e.g., Vanzella et al. 2009 and their Fig. 12).
Moreover, source 859 shows an asymmetric line profile toward the red wavelengths
(Figure~\ref{Fig2}), that is typical of this transition at high redshift.
The spectrum of 1730 is slightly shallower (11200s) and noisier than 859 (close to the
edge of the slit), and prevents us from detecting the asymmetric shape, but
the line width and the equivalent width are not consistent with the [O\,{\sc ii}]$\lambda 3727$ 
doublet. 

Therefore we conclude that the two emission lines are $Ly\alpha$ at the same
redshift $6.387 \pm 0.002$.
The striking accordance of the two redshifts may add support to the
hypothesis that these two objects are multiple images of the same
background source.
If confirmed, this could provide further constraints to the lens model and therefore deserves
future investigation and lens remodeling, which is out of the scope of the present work. In
the following we assume that these are two individual objects and look at the properties
of each of them separately.

$\bullet$~{\it Rest frame UV continuum luminosity at 1500\AA:}
As mentioned above the wide spread on the magnifications allow us to identify an interval of
possible luminosities. Given the observed Y105 magnitudes ($\simeq 1500$\AA)
of $26.42\pm 0.11$ for 859 and $26.34\pm 0.16$ for 1730, the two sources have
unlensed luminosities of $L_{1500} \simeq 0.2$~\muref$L^{\star}_{z=6}$,
adopting $L^{\star}_{z=6}$ from Bouwens et al. (2007).
Even in the more conservative case ($\mu > 5$), these are the faintest
spectroscopically confirmed sources at these redshifts
with such a high signal to noise
(Balestra et al. 2013; Bradac et al. 2012; Schenker et al. 2012).

$\bullet$~{\it Equivalent widths and FWHM of the lines:}
The continuum is not detected in the spectra. Therefore,
we derive the continuum
level under the $Ly\alpha$ transition by using the closest HST band not including the line (Y105),
and correcting for the UV slope $\beta$ (see below).
The rest-frame EWs of 859 and 1730 are $45\pm7$\AA~and $32\pm10$\AA, respectively.
These are typical values if compared with
those observed at similar redshifts among Lyman break galaxies or $Ly\alpha$ emitters
(Stark et al. 2011).
The observed FWHM of the lines is also modest, after correcting for the instrumental profile
they are $\lesssim 150 kms^{-1}$.

%%%%%%%%%%%%%%
\begin{figure}
 \epsscale{1.0}
 \plotone{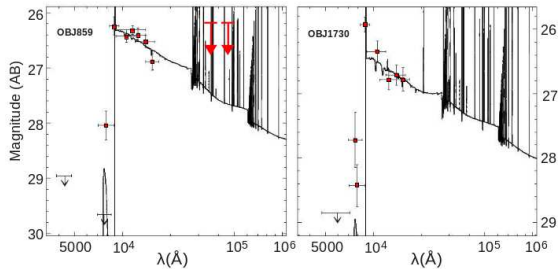} %SED_fit.eps}
\caption{SED fits with BC03 templates are shown. Nebular emission lines are included in the fit.
The two arrows for 859 are 1-sigma lower limits of IRAC 3.6\um~and 4.5\um~channels. \label{Fig3}}
\end{figure}
%%%%%%

$\bullet$~{\it Ultraviolet spectral slope $\beta$ ($F_{\lambda}=\lambda^{\beta}$):}
Following Castellano et al. (2012) and Bouwens et al. (2013), ultraviolet spectral slopes have
been estimated by fitting the near infrared WFC3 magnitudes redward the $Ly\alpha$ line,
using the Y105, J125, F140W and H160 bands (for 859 the F110W band was also available
and has been included in the fit).
Being achromatic, strong lensing is not affecting the colors of the sources.
The measured slopes for 859 and 1730 are very steep, $\beta=-3.02 \pm 0.37$ and
$\beta=-3.01\pm0.56$. Interestingly, this similarity is again consistent with the 
option that these two objects are multiple images.
While the source 1730 is close to a bright galaxy and its photometry has to be taken
with caution, source 859 is isolated and with reliable colors (Figure~\ref{Fig1}).
As noted by Z13, the possible presence of the 2175\AA~dust feature may interfere
with the estimate of the UV slope. However, we tend to exclude this possibility
on the basis of the redshift and the $Ly\alpha$ emission lines that favor
a low dust attenuation.

$\bullet$~{\it Size of the sources:}
As reported in B13, the two sources are resolved in the HST/WFC3 images.
Their isophotal areas (provided by SExtractor) converted into physical units
are 0.8\muref and 0.7\muref sq.kpc.
If they are two distinct objects, we estimate a proper separation of $\sim$ 30 kpc
in the source plane at $z=6.387$, for the range of models described in \S1.

$\bullet$~{\it AGN activity:}
At $z=6.387$ the expected N\,{\sc V}$\lambda1240$ line 
is not detected (see Figure~\ref{Fig2}).
The 1-sigma upper limit N\,{\sc V}$/Ly\alpha$ is $<0.07$ (typical value for AGNs is 10\%, Alexandroff et al. 2013).
Considering the above ratio, that only 5\% of high redshift LAEs are
possible AGNs (Malhotra et al. 2003),
and that they are spatially resolved, we conclude the $Ly\alpha$
emission is due to star formation activity.

%%%%%%%%%%%%%%%%%%%%%%%%
\begin{deluxetable}{lll}
\tabletypesize{\scriptsize}
\tablecaption{Observed and physical parameters for 859 and 1730. \label{physical}}
\tablewidth{0pt}
\tablehead{
\colhead{Quantity} & \colhead{macs0717\_0859} & \colhead{macs0717\_1730}}
\startdata
R.A. (J2000)                     &07:17:38.18 & +37:45:16.9 \\
Decl. (J2000)                    &07:17:37.85 & +37:44:33.7 \\
Redshift                         &  6.387($\pm 0.002$)  & 6.387($\pm0.003$)\\
Y105(observed)                   &  26.42($\pm 0.11$)   & 26.34($\pm0.16$)\\
H160(observed)                   &  26.88($\pm0.15$)    & 26.78($\pm0.18$)\\
H160(unlensed)                   &  28.63+2.5$Log_{10}(\mu_{5})$ & 28.53+2.5$Log_{10}(\mu_{5})$\\
\tableline
$\beta_{UV}$                      &  -3.02($\pm0.37$)    & -3.01($\pm0.56$)\\
$SFR_{UV} (M_{\odot}/yr)$           & 1.6 [1-3] \muref      & 2 [1-5] \muref\\
$SFR_{Ly\alpha} (M_{\odot}/yr)$      & 1.2 [1.0-1.4]\muref    & 1 [0.8-1.2] \muref \\
$M_{\star} (\times 10^{7}M_{\odot})$ & 4 [2-12] \muref       &  2 [2-22] \muref \\
E(B-V)                           & 0.0 [0.0-0.06]      & 0.0 [0.0-0.1] \\
age (Myr)                        & 25 [10-100]         & 10 [10-250]\\
Met. ($Z/Z_{\odot}$)               & 0.02                & 0.2  \\
M1500                            & -18.63+2.5$Log_{10}(\mu_{5})$ & -18.70+2.5$Log_{10}(\mu_{5})$ \\
$\times L^{\star}_{z=6}$            & 0.22 \muref          &0.24 \muref \\
{\it Area} (sq.kpc)              & 0.8 \muref           & 0.7 \muref\\
fwhm($Ly\alpha$)($km/s$)      &100 [70-130]            & 140 [100-180]\\
$EW_{rest}$($Ly\alpha$)(\AA)       &45 [38-53]           & 32 [22-42] \\
flux($Ly\alpha$)$\times 10^{-18}$ &2.8 [2.6-3.0] \muref        & 2.0 [1.8-2.3] \muref\\
\tableline
\enddata
\tablecomments{$Ly\alpha$ fluxes are in units of $ergs^{-1}cm^{-2}$. Physical
properties refer to BC03 models with nebular emission and the associated 68\% 
intervals (in parentheses) correspond to models with $\chi^{2}$ probabilities higher than 0.68.
The SFR($Ly\alpha$) has been derived adopting the Kennicutt (1998) conversion.
Quantities related to $Ly\alpha$ do not include possible
IGM absorption. $\mu_5=1$ corresponds to $\mu=5$.}
\end{deluxetable}
%%%%%%%%%%%%%%%%

\section{Discussion and Conclusions}
As described above, the two discovered sources (or a single one in the
case of multiple images) are the faintest galaxies at $z>6$ ever observed with a
spectroscopic redshift confirmation and well
detected $Ly\alpha$ lines and SEDs. The investigation of {\it new} luminosity regimes
through the strong-lensing magnification gives the opportunity to explore possible new
physical conditions.

\subsection{Nature of the stellar populations}
We examine their rest--frame properties through a SED analysis.
We first derive physical parameters assuming ordinary stellar
populations, i.e.
by comparing the observed SED with a set of Bruzual \& Charlot (2003) templates (BC03),
assuming Salpeter IMF, metallicity of 0.02, 0.2, 1.0 $Z/Z_{\odot}$,
and E(B-V) spanning the range [0.0 -- 1.0]. The current 1-$\sigma$ lower limits from
IRAC (3.6\um~and 4.5\um~channels) for 859 are $\simeq 26.1$AB, too shallow to
provide solid constraints on [O\,{\sc iii}]$\lambda 5007$+$H\beta$ and $H\alpha$ nebular emissions.
The other source 1730 is contaminated by close brighter galaxies.
The SED fitting with BC03 includes nebular line and continuum emission following
Schaerer \& de Barros (2009) (see Castellano et al. 2014 for further details).
The output of this exercise is listed in Table~\ref{physical}.
Regardless of the adopted $\mu$, the two sources turn out to be very
small ($\lesssim$ 1 sq.kpc), with low SFRs ($\simeq 1-2 M_{\odot}/yr$) and
low stellar masses of $< 10^{8} M_{\odot}$.
The properties related to colors (i.e., independent from the magnification $\mu$),
such as dust attenuation, age, and metallicity, are consistent with newborn objects. 
Adopting the standard Kennicutt conversions (Kennicutt 1998) 
and correcting for the IGM attenuation of $Ly\alpha$ photons 
(e.g., $>50$\%, Dijkstra \& Jeeson-Daniel 2013),
we obtain SFR($Ly\alpha$)$\gtrsim$SFR(UV), where SFR(UV) is derived
from the SED fit. This is indicative of
ages $<100$Myr, E(B-V)$\simeq0$ (Verhamme et al. 2008) 
and $f_{esc}$($Ly\alpha$) close to unity (Atek et al. 2008, 2013).

The SEDs can be reproduced with ordinary stellar populations, 
albeit the best solutions typically lie close to the edge of the 
parameter space (e.g., Z, age and E(B-V)). Fixing 
$Z=Z_{\odot}$, the resulting ages are forced to the minimum value, $10$Myr.
For this reason it is interesting to extend the investigation toward
a possible presence of
younger and/or extremely metal poor (EMP, $Z\simeq 1/2000 Z_{\odot}$) and 
PopIII stars ($Z=0$). For this purpose we consider the SED fitting
and the predicted HST/WFC3 colors provided by 
Raiter et al. (2010), Inoue et al. (2011) and Zackrisson et al. (2013),
(R10, I11 and Z13, respectively), that also include nebular contribution.
The observed UV slope is compatible either with
very young, but still standard (PopII) stellar populations (BC03), or with
EMP/PopIII stars. In particular, 859 (with the most reliable photometry)
has a $\beta = -3.02\pm0.37$ that is consistent with an age $\simeq 1$Myr
if $Log_{10}(Z/Z_{\odot})=0$ or an age $\simeq 1-100$Myr if
$Log_{10}(Z/Z_{\odot})<-4$ (as shown in I11, Fig. 11). \footnote{We note that the 
probability to observe a galaxy of a few Myr old is generally small,
because of its short time.}
Similarly, compared with the models of Z13, the UV slope
is compatible with metal poor stars if ages are $>10$Myr, and
even PopIII if compared with R10 (assuming we are observing the stellar 
component). Conversely, the observed $Ly\alpha$ EWs would suggests that
we are dealing with standard stellar populations, given that PopIII stars 
are often associated with $Ly\alpha~EW \sim 500-1500$\AA~rest-frame 
(Schaerer et al. 2003; R10; I11). A large IGM attenuation of the $Ly\alpha$ 
line ($>90$\%) could hide a intrinsic EW$>500$\AA, making it still compatible 
with the PopIII interpretation. However the influence of the IGM is highly 
uncertain here (see also Laursen et al. 2011; Dayal et al. 2011; Dijkstra \& Jeeson-Daniel 2013).
Another possibility is that the $Ly\alpha$ EW could be lowered for extremely
metal poor ($Z/Z_{\odot}<10^{-4}$) and even PopIII ($Z/Z_{\odot} = 0$) galaxies
if $f_{esc}>0$ (Z13).
For example, I11 found a $Ly\alpha$ EW of $\simeq 65$\AA~for $Z=0$ and 10Myr
constant star formation, when $f_{esc} = 0.9$.

While it is hard to make definitive statements about the populations content of 
these $z\sim6.5$ sub-luminous galaxies given the current information we have 
about them, we observe that they are
overall less evolved than their more massive counterparts and
their very blue UV colors could be explained even without having to invoke PopIII stars,
although we certainly cannot exclude their presence in the stellar populations 
(e.g. Finkelstein et al. 2010). 
These kind of galaxies could be 
examples of very low chemical enrichment, dust-free  systems, barely higher than 
the pristine gas that is probably still feeding their activity of star formation.

\subsection{Cosmic reionization}
Regardless of the nature of the stellar populations, the potential role these
sources have in the framework of cosmic reionization is intriguing.
It is believed that the abundant, fainter galaxies ($M_{\star}<10^{9} M_{\odot}$)
could significantly contribute to, or even be the
dominant populations in, providing the ionizing radiation
(e.g., Fontanot et al. 2013; Ferrara \& Loeb 2013; Yajima et al. 2011;
Razoumov \& Sommer-Larsen 2010; but see Gnedin et al. 2008).

The direct measure of the escape fraction of ionizing radiation ($f_{esc}$) is
in principle possible at $z<4$. At higher redshifts the measurement is
unfeasible due to the complete IGM attenuation of the Lyman continuum.
Nonetheless it is worthwhile to investigate each of the main components that build
the $f_{esc}$ quantity.
As discussed in Vanzella et al. (2012), the $f_{esc}$ parameter is the product of
the gas transmission $T_{900}^{HI}=exp(-\tau_{900})$ and the dust transmission
$T_{dust}=10^{-0.4 \times A900}$.
To first order, Lyman continuum
emitters should have both low dust content $A900$ {\em and} low optical depth
$\tau_{900}$.
Interestingly, the two sources described in this letter could match such requirements.
First, given the very steep UV continuum the term related to dust attenuation is
significantly higher than zero 
($T_{dust}=10^{-0.4 \times A900} \simeq 1$, as reported in Siana et al. (2007)
by extrapolating the Calzetti extinction law down to the Lyman continuum, 
$A1500=A900=0$ if E(B-V)=0).
Second, even though addressing the gas attenuation in the interstellar medium
with current data is admittedly less reliable, it is worth noting that
the presence of $Ly\alpha$ emission would not be in contrast with a $f_{esc}>0$.
As discussed in Nakajima \& Ouchi (2013), the EW($Ly\alpha$) remains 
almost unchanged if $f_{esc}$ is $\simeq 0 - 0.8$. 
If the IGM is attenuating $<70$\% of the line 
(Dijkstra \& Jeeson-Daniel 2013), the resulting intrinsic FWHM  
($\lesssim 300 kms^{-1}$)
would be in line with possible low HI column density in front of the 
stars (Schaerer et al. 2011).
The reason is that $Ly\alpha$ resonance scattering is less effective if $N_{HI}$ is low,
and photons escape easily along the shorter path, decreasing the FWHM.
Moreover, the observed UV slope $\beta \simeq -3$ could also indicate a
deficit of nebular continuum, allowing the stellar component to emerge in the 
observed SED (i.e., $f_{esc}>0.5$, R10, I11, Z13).
This, could, in turn, be the telltale of efficient feedback in these
systems, capable of either sweeping away or ionize a significant fraction of the gas
surrounding the stars, a mechanism advocated by theoretical models in low--mass
halos to self-regulate star formation. As a consequence, a proportionally higher
fraction of ionizing radiation could be leaking out of these systems ($f_{esc}>0$)
compared to their more massive counterparts and be available to keep the IGM ionized.

Regarding the single or multiple nature of the sources,
the similarity in the physical and observed characteristic we derived in this work,
would support they are multiple images of a single background $z=6.4$ galaxy, but the 
different mass models we examined remain inconclusive regarding this option.

As discussed in Zackrsson et al. (2012) and Z13, such galaxies represent the
ideal candidates for future near- and
mid-infrared spectroscopic observations, especially in the investigation of
the interplay between the UV slopes and the equivalent width of $H\beta$ lines, and
its relation to the $f_{esc}$ parameter. Future facilities such as JWST and extremely
large telescopes will address this issues.

\acknowledgments
We acknowledge the support from the LBT-Italian
Coordination Facility for the execution of observations, data
distribution and reduction. We thank Gianni Zamorani, Marco Mignoli and
Francesco Calura for useful discussions. This work utilizes gravitational lensing
models produced by PIs Bradac, Ebeling, Zitrin \& Merten, Sharon, and
Williams funded as part of the HST Frontier Fields program conducted
by STScI. STScI is operated by the Association of Universities for
Research in Astronomy, Inc. under NASA contract NAS 5-26555. The lens
models were obtained from the Mikulski Archive for Space Telescopes
(MAST). Support for AZ is provided by NASA through Hubble Fellowship 
grant \#HST-HF-51334.01-A awarded by STScI.
AF acknowledges the contribution of the EC FP7 SPACE project ASTRODEEP (Ref.No: 312725).
LI is partially supported by CATA-Basal, Conicyt. We acknowledges financial
contribution from PRIN-INAF-2010 and PRIN-INAF-2012.

\end{document}